\documentclass[prl,twocolumn,superscriptaddress]{revtex4}
\usepackage{graphicx}
\usepackage{epsfig}

\usepackage{amssymb,amsmath,amsfonts,hyperref}
\usepackage{wasysym}
\usepackage{latexsym}

\newcommand{\be}{\begin{eqnarray}}
\newcommand{\ee}{\end{eqnarray}}

\begin{document}
\preprint{APS/123-QED}
\title{Semiclassical spin liquid state of easy axis Kagome antiferromagnets}
\author{Arnab Sen}
\address{{\small Department of Theoretical Physics,
Tata Institute of Fundamental Research,
Homi Bhabha Road, Mumbai 400005, India}}
\author{Kedar Damle}
\address{{\small Department of Theoretical Physics,
Tata Institute of Fundamental Research,
Homi Bhabha Road, Mumbai 400005, India}}
\address{{\small Physics Department, Indian Institute of Technology Bombay, Mumbai 400076, India}}
\author{R.~Moessner}
\address{{\small Max-Planck-Institut f\"ur Physik komplexer Systeme, 01187 Dresden, Germany}}
\vspace{-2 cm}
\date{\today}
\begin{abstract}
Motivated by recent experiments on Nd-langasite, we consider the effect of
strong easy axis single-ion anisotropy $D$ on $S > 3/2$ spins interacting
with antiferromagnetic exchange $J$ on the Kagome lattice. When $T \ll
DS^2$, the collinear low energy states selected by the anisotropy
map on to configurations of the classical Kagome lattice Ising
antiferromagnet. However, the low temperature limit is
quite different from the cooperative Ising paramagnet that obtains
classically for $T \ll JS^2$. We find that sub-leading ${\mathcal O}(J^3S/D^2)$ multi-spin
interactions arising from the transverse quantum dynamics result in
a crossover from an intermediate temperature classical
cooperative Ising
paramagnet to a semiclassical spin liquid with distinct short-ranged
correlations for $T \ll J^3S/D^2$. 
 
\end{abstract}

\pacs{74.20.Mn 74.30.+h 74.20.-z 71.55.Jv}
\vskip2pc

\maketitle

\begin{figure}
\includegraphics[width=\hsize]{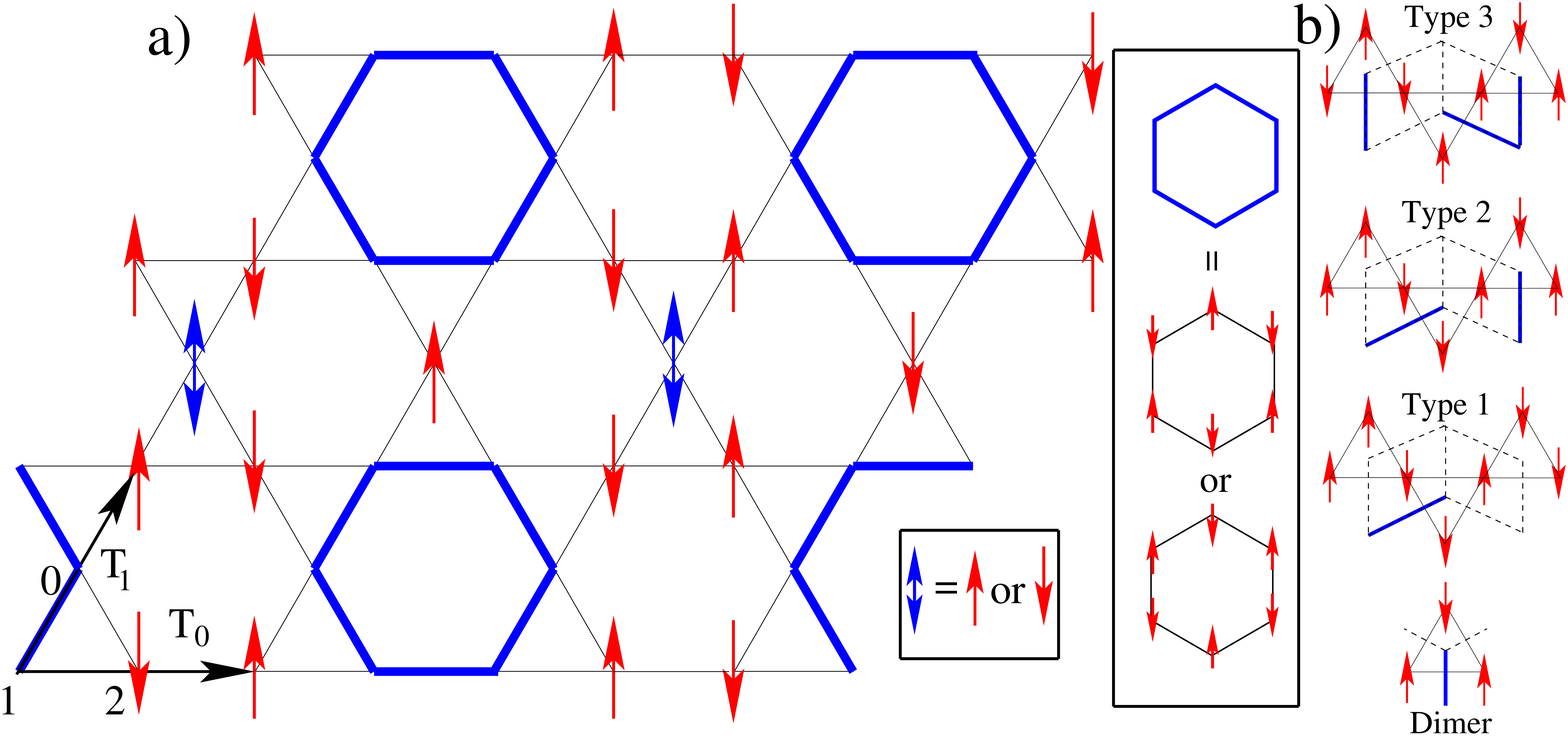}
      \caption{(color online). a) A $L_x=4$, $L_y=3$ piece of the Kagome
      lattice, with principal directions $T_0$ and $T_1$, and three sublattices
      of sites marked; also shown is a class of partially ordered states with
      macroscopic entropy that simultaneously minimize $J_1$ and $J_2$. b) Unfrustrated bonds of
      type $n$ and mapping of Ising spins to dimers on the
      dual dice lattice.}
      \label{Fig1}
  \end{figure}

{\em Introduction:}  The insulating magnet Nd-langasite has localized
Nd$^{3+}$ moments that live on sites of a $2$-dimensional Kagome lattice
made up of corner-sharing triangles (Fig~\ref{Fig1}a). These moments carry
a total angular momentum  ${\bf
  J}_{{\mathrm{ion}}} =9/2$  and are subject to a strong single-ion anisotropy term $D \sim 10
K$~\cite{Zorko_etal_PRL2008} that picks out the crystallographic $c$ axis as the
common easy axis of all the spins~\cite{Langasite1}. Although they
interact with a sizeable nearest neighbour antiferromagnetic
exchange coupling $J \sim 1.5 K$~\cite{Zorko_etal_PRL2008} (corresponding
to a Curie-Weiss temperature $\Theta_{CW} = -52$K), Nd-langasite does not 
exhibit any magnetic order
down to $50$ mK~\cite{Langasite_neutrons1,Langasite_neutrons2,Zorko_etal_PRL2008}. 
Such behaviour contrasts strikingly 
to conventional  insulating solids with localized magnetic moments and concomitant 
short-ranged exchange interactions~\cite{Goodenough}---these 
usually enter a magnetically
ordered state at low temperature, the nature of which can often be
understood quite simply in terms of the classical energetics of the leading
exchange interactions. 

When these interactions compete due to the geometry of the
lattice, as is the case in Nd-langasite, one often obtains a large degeneracy
of inequivalent classical ground states which prevents ordering and instead 
results in unusual cooperative
paramagnetic behaviour for a range of intermediate temperatures below $\Theta_{CW}$ 
-- such systems are called frustrated magnets~\cite{Moessner_CanJP}. 
Frequently, quantum effects and sub-leading
interactions eventually do lead to an (often complex) ordered state at
still lower temperature---Kagome lattice magnets provide many examples of this~\cite{Kagomeexpts1,Kagomeexpts2,Kagomeexpts3}. 
In some other cases, such as the $S=1/2$ Kagome
magnet {\em
  herbertsmithite}~\cite{Helton_etal_PRL2007}, there is apparently no tendency of the spins to form an ordered
arrangement even at the lowest temperatures accessible to experiment.

Systems such as Nd-langasite and Herbertsmithite provide possible
realizations of so-called spin-liquid states, which have been the
subject of sustained theoretical activity~\cite{Misguich_Lhuiller_review} going
back to the seminal work of Fazekas and Anderson~\cite{Fazekas_Anderson}.
Much of this theoretical activity has focused on the challenging case
of Heisenberg exchange interactions (isotropic in spin space) and low
spin $S=1/2$ (in which quantum fluctuations are expected to be
strongest)~\cite{Misguich_Lhuiller_review}. In the opposite Ising limit in
which the exchange interactions only couple one component of neighbouring
$S=1/2$ moments on the Kagome lattice, there are no quantum effects and the
system remains in a cooperative paramagnetic state with short ranged spin correlations
all the way down to zero temperature~\cite{BookLiebmann}.

In order to model the case at hand, Nd-langasite, 
we consider  moments with
larger spin $S >  3/2$ interacting with isotropic Heisenberg exchange $J$ on the
Kagome lattice, but which---unlike the case considered in
\cite{Tchernyshyov}---are subjected to a strong single-ion anisotropy
$D$ that picks out a common easy axis for all the moments:  
\be
H = J\sum_{\langle ij \rangle} \vec{S}_i \cdot \vec{S}_j -D\sum_i (S^z_i)^2
\ee
When $D$ dominates over $J$,  the moments prefer collinear spin
states that correspond to configurations of a classical Ising
antiferromagnet. However, quantum fluctuations induced by transverse components
of the exchange coupling can  lead to low temperature behaviour
quite different from the classical cooperative Ising paramagnet, and here we explore this possibility in some detail.

Our results are readily stated: We find that such easy axis magnets {\em do
  not}  develop any long-range order down to very low temperature, and are thus
  good examples of genuine spin-liquid behaviour in a system of quantum spins.
More precisely, we find that there are two qualitatively distinct regimes
  separated by a cross-over temperature $T^{*} \sim J^3S/D^2$. For $T^{*} <  T
  \ll JS^2$, the short ranged spin correlations of the system are
  well-described by the correlations of the cooperative Ising paramagnet 
described above. Below
  $T^{*}$, the leading effects of {\em virtual} quantum fluctuations begin to
  dominate, leading to a qualitatively different semiclassical spin liquid
  regime in which the liquid structure factor of the spins
  encodes distinctive short-ranged correlations, but there is no long range
  order of any kind (this is in sharp contrast to easy axis Kagome
antiferromagnets with $S=1$ moments, where the quantum dynamics connects
different classical ground states and induces spin nematic 
order~\cite{Damle_Senthil}).

{\em Effective Hamiltonian:} When $D$ dominates over $J$, the system prefers
{\em collinear spin configurations} with $S^z_i = \sigma_i S$ (with $\sigma_i =
\pm 1$). This degenerate ground-state manifold of the unperturbed problem can
be thought of in terms of configurations of the Ising pseudo-spin variables
$\sigma$. The low energy physics in this regime is then best described in terms
of an effective Hamiltonian $\mathcal{H}$ that encodes the splitting of this degenerate Ising
subspace to each order in $J/D$. To order $J^3/D^2$, $\mathcal{H}$ is given 
as~\cite{Sen_Damle_Vishwanath} 
\be
\mathcal{H}_{\mathrm{eff}} &=& J_1 \sum_{\langle ij \rangle}\sigma_i \sigma_j -
J_2 \sum_{\langle ij \rangle} \frac{1-\sigma_i
  \sigma_j}{2}(\sigma_i H_i + \sigma_j H_j) 
\ee
where $J_1 = JS^2$, $J_2 = \frac{S^3J^3}{4D^2(2S-1)^2}$, and the {\em exchange
  field} $H_i \equiv \Gamma_{ij}\sigma_j$ with $\Gamma_{ij}=1$ for nearest
neighbours and zero otherwise. In the above, the first term corresponds to the
leading effect of the $z$ component of the spin exchange, while the second term
arises from  {\em virtual} quantum transitions of pairs of
anti-aligned spins out of the low energy Ising subspace. An additional
  ${\mathcal O} (J^{2S}/D^{2S-1})$ pseudo-spin exchange term, representing 
  real quantum transitions, is subleading for $S >
  3/2$~\cite{Sen_Damle_Vishwanath}.
  
{\em Classical Ising regime:} For temperatures $T$ well above the exchange energy scale $JS^2$, but lower
than the anisotropy energy $DS^2$, the behaviour of the system will be that of
to the high temperature paramagnetic regime of the classical Ising model on the
Kagome lattice. As the temperature is lowered, the exchange
 begins to make itself felt, and the system crosses over to an
intermediate temperature regime $J_2 \ll T \ll J_1$ whose physics is controlled
by the ground states of the classical Kagome lattice Ising antiferromagnet
(KIAF). 

In these ground states each triangle has exactly one frustrated bond
(connecting a pair of aligned spins). This 'minimally frustrated' ensemble of states has a residual
entropy of $0.502k_B$ per site~\cite{BookLiebmann}, and the loss of entropy
coming from the high-temperature paramagnet is reflected in a featureless peak in the
specific heat $C_v$ (the area under the $C_v/T$ curve measuring
this entropy loss). 

This  ensemble of ground states can be conveniently represented
by dimer coverings on the dual dice lattice wherein a dimer is placed on every
link of the dual lattice that intersects a frustrated Kagome lattice bond
(Fig~\ref{Fig1}b); note that in this dimer model, a hard dimer constraint is
operative {\em only} on the $3$-coordinated dice lattice sites, while the $6$-coordinated sites have a soft constraint that an even number of dimers touch them. 

As mentioned earlier, this ensemble~\cite{BookLiebmann} 
yields
correlations that are extremely short-ranged and hence a 
featureless spin structure factor (Fig~\ref{Fig3}a). The temperature regime $J_2
\ll T \ll J_1$  is thus a {\em classical cooperative Ising paramagnet}
with a
reduced (compared to the high temperature Ising paramagnet) value of the
magnetization fluctuations as reflected in  $T\chi$ ($\chi$ being
the linear susceptibility to a magnetic field along the easy axis).
\begin{figure}
\includegraphics[width=\hsize]{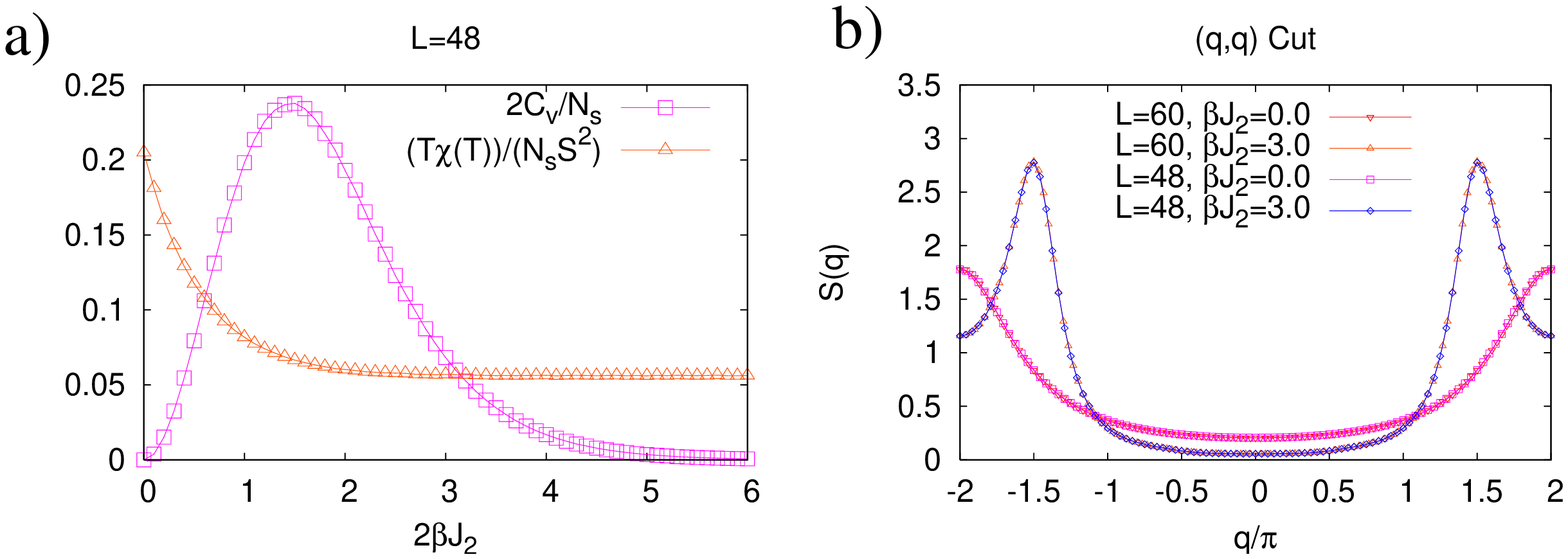}
      \caption{(color online). (a)Specific heat $C_v$ and uniform
susceptibility $\chi$ in the crossover region. (b) The $q_x=q_y$ cut for $S(\vec{q})$ is also
      shown for two different system sizes ($L=48,60$) for two different
      temperatures ($\beta J_2=0.0,3.0$).}
      \label{Fig2}
  \end{figure}

{\em Minimizing $J_2$:} To understand the behaviour of the system at
still
lower temperatures, we first focus on the $T=0$
ground states of $\mathcal{H}_{\mathrm{eff}}$ and show that minimally frustrated configurations which also minimise
the multispin interaction $J_2$ satisfy the criterion that {\em no spin be the
  minority spin of both triangles to which it belongs}. To this end,
we
first note that the multispin
interaction, projected to the minimally frustrated Ising subspace, assigns a
`potential energy' $-2J_2 (n-1)$ to an unfrustrated bond of type $n$---here
we classify an unfrustrated bond to be of type $3$ ($1)$ if the two spins it
connects are both majority (minority) spins of the {\em other} triangles to
which they belong, while an unfrustrated bond connecting one majority and one
minority spin is of type $1$ (Fig~\ref{Fig1}b).

As an unfrustrated type $n$ bond of a minimally frustrated
Ising configuration is surrounded by a double rhombus of the
dual dice lattice with $n$ dimers on
its perimeter (Fig~\ref{Fig1}b), the projected multispin interaction $J_2$
translates to a dice-lattice dimer model with 
interactions on double-rhombii surrounding unfrustrated bonds
\be
H_D = -4N_{t}J_2 \sum_{n=1}^{3}(n-1)f_n
\ee
 where $N_t$ is the number of triangles in the kagome lattice and $f_n$
is the  fraction of unfrustrated bonds of type $n$, i.e with $n$ dimers on the
perimeter of the corresponding double-plaquette (Fig~\ref{Fig1}b.).

Thus, up to a constant, the interaction energy counts the {\em total number} of
dimers on the perimeters of all double-plaquettes corresponding to unfrustrated
bonds. 
 Since every single plaquette with $n$ dimers on its perimeter is
 part of $4-n$ such double-plaquettes, 
 the $n$ dimers will contribute with multiplicity $4-n$ to this
 total number. Thus, if the fraction of single plaquettes with $n$
 dimers on
their perimeter
 is $g_n$ (with $\sum_{n=0}^{2} g_n = 1$, since each elementary plaquette can
 have $0$, $1$, or $2$ dimers on its perimeter), the interaction
 energy of the configuration is  $-N_{P} \sum_{n=0}^{2} n(4-n)g_n$, where
 $N_{P} = 3N_t/2$ is the number of elementary plaquettes. The interaction term
 thus assigns energies---quadratic in the number of dimers---to different
 types of elementary plaquettes of a dice lattice dimer configuration: 
\be
H_D = 2J_2\sum_{P} n^2|n P\rangle \langle nP|
\ee 
where $|nP\rangle$ denotes elementary plaquettes with $n$ dimers on their
 perimeter. Since each dimer is on the perimeter of two elementary plaquettes, we have $\sum_{n=0}^{2}ng_n = 4/3$, and this constraint along 
with $\sum_{n=0}^{2}g_n =1$ allows us to minimize the potential energy $H_D$: $H_D$ is minimized when $g_0=0$ (which fixes $g_1=2/3$ and $g_2=1/3$) and this immediately gives the criterion that no spin be the minority spin of both the triangles to which it belongs.

We have investigated the set of minimally frustrated configurations that satisfy this minority spin rule in some detail, and find that it is possible to construct a large subset of states
(i.e. with macroscopic entropy) satisfying this rule, and related to
each other by local spin flips (Fig~\ref{Fig1}a). This construction
immediately provides a lower bound of $k_B \ln(2)/6$ per site on the
entropy of the ground states of
$H_D$. The question then arises whether the correlations in the
$T \rightarrow 0$ limit remain short-ranged, or whether order by disorder
occurs here, and below we address this question numerically.
\begin{figure}
\includegraphics[width=\hsize]{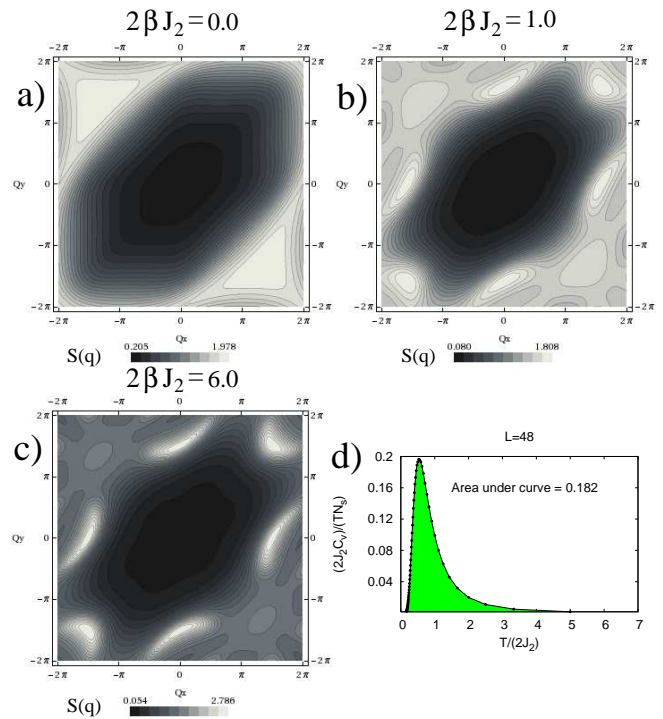}
      \caption{(color online). (a-b-c)Spin structure factor $S(\vec{q})$ shown for
      three different temperatures, showing crossover to semiclassical
spin liquid regime. d) Area
      under the $C_v/T$ curve gives the reduction in entropy associated with
      the crossover to the semiclassical spin liquid.  }
      \label{Fig3}
  \end{figure}

 {\em Loop algorithm:} In order to obtain reliable
numerical results that can be used to settle this delicate question,
we have
developed a new loop algorithm that can handle the non-trivial
Boltzmann
weight associated with $H_D$ as well as keep track of the different
constraints on $3$- and $6$-coordinated sites, while retaining
the efficiency of the usual hard-core dimer model loop
algorithms~\cite{Sandvik_Moessner,Alet_etal}. 

The algorithm proceeds as follows:
A change in dimer configuration is initiated by starting
at a randomly chosen $6$-coordinated site (say $\mu_0$) and moving, with equal
probability, to one of its
six $3$-coordinated neighbours, say $i$.
If the link $\langle \mu_0 i \rangle$ is covered by a dimer, this
dimer is rotated about
the {\em pivot} site $i$, so that it now covers link
$\langle \mu i \rangle$---$\mu$ is chosen from the three
possible $6$-coordinated neighbours of $i$ according to a table of
probabilities satisfying detailed balance (if $\mu = \mu_0$,
the attempted update of the dimer configuration ends without making
any change in the configuration).
On the other hand, if the link $\langle \mu_0 i \rangle$ is
unoccupied by a dimer, and the dimer touching $i$ covers a different link
$\langle \mu i \rangle$, this dimer is rotated about
pivot $i$ to cover link $\langle \mu_0 i \rangle$ with
a certain probability $p$ chosen to satisfy detailed balance
(conversely, with probability $1-p$, the attempted update ends without
making any change in the dimer configuration).

In both cases above, the first step of the update procedure 
results in a violation of the (soft) constraint
at $\mu_0$ as well as at another
$6$-coordinated
site $\mu$. By repeating this set of moves with probabilities chosen
to satisfy detailed balance at each step, the site $\mu$
can be moved around in a closed loop until
it finally meets $\mu_0$ again and heals
the `defects' that were originally introduced at that both sites.
When this happens, a large change in the dimer configuration is
effected
along a closed loop of links, and this large change can be
accepted with unit probability, thereby providing an efficient
means of sampling the Gibbs distribution associated with $H_D$. 

{\em The semiclassical spin liquid:}
From numerical simulations on $L_x=L_y=L$ size systems using this algorithm
(with $L$ ranging from $10$ to $60$), we see no evidence at all of any  phase transition as we
lower the temperature to access the $T \rightarrow 0$ limit. This is evident
from the behaviour of the specific heat per site, which converges
very quickly with system size, and does not show any
singularity in the thermodynamic limit (Fig~\ref{Fig2}a). In addition, the spin-spin
correlators as well as the bond-energy correlators show no long range
order at any wavevector down to the lowest temperatures we study (Fig~\ref{Fig2}b). The
system
thus remains in a short-ranged ordered spin-liquid state down to the
lowest temperatures.

Although there is no phase transition, we find that the liquid state
at low temperature is quite different from the intermediate
temperature
cooperative Ising paramagnet. This {\em crossover} to a distinct
semiclassical spin liquid regime
at low temperature (Fig~\ref{Fig2}a) is evident for instance in the temperature
dependence of the specific heat per site $C_v/N_s$: The $C_v$ vs $T$
curve shows a distinct but non-singular peak at $T^{*} \approx 1.3 J_2$ that
reflects the loss of entropy during this crossover from the cooperative Ising
paramagnet to the low temperature limit in which the configurations sampled predominantly obey
the minimum $J_2$ constraint (from the area under the $C_v/T$ curve
(Fig~\ref{Fig3}d) and knowledge of the residual entropy of the cooperative Ising
paramagnet, we estimate the residual entropy of the
semiclassical spin liquid to be $0.32 k_B$).

A clear signature of this crossover to the semiclassical spin liquid regime
below $T^{*}$ can be obtained by monitoring the spin structure factor
(that can be probed in neutron scattering experiments)
$S(\vec{q}) =
|S_0(\vec{q})\exp(iq_y/2)+S_1(\vec{q})+S_2(\vec{q})\exp(iq_x/2)|^2$
where $S_{\alpha}(\vec{q})$ is the Fourier transform of the spin
density on sublattice $\alpha$ of the Kagome lattice 
and $q_x$ ($q_y$) refers to the
projection of $\vec{q}$ on to lattice direction $T_0$ ($T_1$) measured
in units of inverse Bravais lattice spacing (Fig~\ref{Fig1}a).
From Fig~\ref{Fig3} a), b), c), we see that the structure factor
evolves continuously from being quite featureless in the classical cooperative 
Ising regime
$T^{*} \ll T \ll J_1$ to developing characteristic crescents of high intensity
diffuse scattering in the low temperature semiclassical spin liquid regime $T
\ll T^{*}$, with precursors of these features being already present at $T \sim
2T^{*}$. In addition, this crossover is also characterized by a change in the
magnetization fluctuations as reflected in the value of $T \chi$ (Fig~\ref{Fig2}a).   

{\em Experiments on Nd-langasite:}
As mentioned earlier, recent experiments on the spin-$9/2$ easy axis Kagome
antiferromagnet Nd-Langasite have seen a liquid-like state with fluctuating
moments and no long range order down to $50mK$. From the estimated
values~\cite{Zorko_etal_PRL2008} of the isotropic exchange interaction and the single-ion
anisotropy ($J \sim 1.5K$, $D \sim 10K$), the
crossover from a classical cooperative Ising paramagnet to a semiclassical spin liquid
only occurs for temperatures significantly below $50mK$ ($T^{*} \approx 16 mK$) in this model
with an isotropic exchange and a simple single-ion anisotropy term (which
is expected~\cite{Zorko_etal_PRL2008} to be a good starting point for Nd-Langasite).

We therefore expect a simple classical Ising description to work fairly well
for the bulk of the temperature range studied in the recent low temperature
experiments. Since correlations in the cooperative Ising paramagnet are extremely
short-ranged and featureless, this is consistent with the fact that data from
recent neutron scattering experiments (which probe the spin structure factor)
show nearly featureless diffuse scattering that can be fit quite well to a
model~\cite{Langasite_neutrons1,Langasite_neutrons2} of spin correlations in
which nearest neighbour spins are correlated, but there are no correlations of
spins further away from each other.

We hope that our results provide further motivation to study this magnet at
still lower temperatures, at which the crossover to the interesting low
temperature semiclassical spin liquid should become apparent. Another
possible avenue for exploring this crossover in greater detail involves
identification of other easy axis Kagome antiferromagnets in which the
separation of scales between $J$ and $D$ is not so large (or the overall scale
of both $J$ and $D$ is somewhat larger) so that the crossover to the
semiclassical spin liquid occurs at more easily accessible temperatures and we hope that the results of our work provide motivation for
exploring this possibility as well.


{\em Acknowledgements:} We thank F.~Bert and D.~Dhar for useful discussion and
correspondence, and A.~Vishwanath for collaboration on related work. We
acknowledge computational resources at TIFR and support from DST
SR/S2/RJN-25/2006 (KD).

\end{document}